\documentstyle[multicol,aps,prb,epsf]{revtex}
\begin{document}
\pagestyle{empty}
\newcommand{\bc}{\begin{center}}
\newcommand{\ec}{\end{center}}
\newcommand{\be}{\begin{equation}}
\newcommand{\ee}{\end{equation}}
\newcommand{\beqn}{\begin{eqnarray}}
\newcommand{\eeqn}{\end{eqnarray}}

\begin{multicols}{2}
\narrowtext
\parskip=0cm

\noindent
{\large\bf Quiet Violin,
Comment on "The dead zone for string players"}  
\smallskip

I read with interest your recent article about how a violin generates sound, 
which reported that Johan Broomfield and Michael Leask had found it becomes 
impossible to produce a sound from a string that is bowed at its midpoint 
(January p5). However, this finding is trivial if one knows the mechanism of 
bowing

When a horse-tail bow is drawn across a string, tiny microscopic "nails" on 
the surface of the bow constantly pluck the string as the bow moves over the 
string (see, for example, E G Gray, Splitting Hairs, Strad vol.82 p107-108 
(1989)). 
This is different from what happens when you pluck a string with your finger. 
The string can then vibrate freely once your finger has left the string. 
But with a bow, the nails continuously intervene in the vibration, producing 
a sort of "forced" or "kicked" vibration. Once one nail has passed over the 
string, the string can only vibrate freely for a short period before the next 
nail arrives. Indeed, the nails are so closely spaced that it is perhaps 
better to describe them as "brushes" or "combs.

It is therefore better for the amplitude of vibration at the bowing point to 
be about the same as the spacing of the nails so that the intervention of the 
nails is minimized. If the amplitude is larger than the spacing of the nails, 
the string does not manage to vibrate much before the next nail arrives. 
(Actually, the presence of the next nail does not stop the vibration 
completely; it only "constrains" the vibration to the position of the next 
nail.) This mechanism applies to every part of the string, but the influence 
of the next nail becomes most significant as you bow at the 
midpoint --***because the amplitude of vibration is then at its largest***.  


\bigskip
\noindent
{Julian Juhi-Lian Ting}

{\small \noindent
Taichung, Taiwan, Republic of China \\
jlting@yahoo.com
}

\bigskip
\noindent
Date: \today

\end{multicols}

\end{document}